\documentclass[epj]{webofc}
\usepackage[varg]{txfonts}   
\wocname{EPJ Web of Conferences}

\woctitle{INPC 2013}
\begin{document}
\title{Search for new Physics with the $\pi\rightarrow e\nu$ Decay}


\author{Luca Doria\inst{1}\fnsep\thanks{\email{luca@triumf.ca}} (for the PIENU Collaboration)}

\institute{TRIUMF, 4004 Wesbrook Mall, Vancouver,BC V6J 2A3 (Canada)}

\abstract{%
In the Standard Model, lepton universality refers to the identical electroweak 
gauge interactions among charged leptons. The measurement of the branching ratio
$R_{e / \mu} = \frac{\Gamma(\pi\rightarrow\ e \nu)}{\Gamma(\pi\rightarrow\mu\nu)}$
is one of the most stringent tests of lepton universality between the first two generations. The TRIUMF PIENU experiment
aims at the most precise test of universality measuring $R_{e / \mu}$ with 0.1\% precision. 
The measurement will provide improved constraints to physics beyond the SM or uncover new scenarios if a disagreement is found.
}
\maketitle
\section{Introduction}
\label{intro}
The measurement of the branching ratio
\begin{equation}
R_{e / \mu} = \frac{\Gamma(\pi\rightarrow\ e \nu + \pi\rightarrow\ e \nu \gamma)}{\Gamma(\pi\rightarrow\mu\nu + \pi\rightarrow\mu\nu\gamma )}
\end{equation}
is the most precisely calculated observable involving quarks in the Standard Model (SM) \cite{r0,r1,r2}.
The latest modern calculation \cite{r3} takes into account the pion structure using chiral perturbation theory
to order $O(e^{2}p^{4})$ and radiative corrections as well:
\begin{equation}
R_{th} = (1.2352 \pm 0.0001) \times 10^{-4} .
\end{equation}
The precision comes form the cancellation of strong interaction effects in the ratio and from the appearance of 
structure dependent terms only through electroweak corrections.
The current experimental value \cite{exp1,exp2} is
\begin{equation}
R_{exp} = (1.230 \pm 0.004) \times 10^{-4} .
\end{equation}
The PIENU experiment aims at improving the experimental precision by about a factor 5,
and therefore testing the SM prediction to better than $0.1\%$ level.
The experiment has the potential to uncover physics beyond the SM, measuring a deviation
from the theoretical calculations. 
There are many possible extensions of the SM which can modify $R_{e / \mu}$, like heavy neutrinos, extra dimensions,
leptoquarks, compositeness, charged Higgs bosons or (R-parity violating) SUSY. 
Since the $\pi\rightarrow e\nu$ decay is helicity-suppressed, $R_{e / \mu}$ is very sensitive to helicity-unsuppressed 
couplings like the pseudo-scalar ones. Pseudo-scalar contributions through interference terms are 
proportional to $1/m_{H}^{2}$, where $m_{H}$ is the mass of the hypothetical particle. 
On the contrary, lepton flavor violating  decays such as $\mu\rightarrow e \gamma$ receive $1/m_{H}^{4}$ contributions.
The deviation from the SM prediction can be parameterized as \cite{r0}:
\begin{equation}
1-\frac{R_{exp}}{R_{SM}} \sim \pm \frac{\sqrt{2\pi}}{G_{\mu}} \frac{1}{\Lambda_{PS}}
\frac{m_{\pi}^{2}}{m_{e}(m_{d}+m_{u})} \approx \left( \frac{1 {\rm TeV}}{\Lambda_{PS}} \right)^{2} \times 10^{3}.
\end{equation}
The parameter $\Lambda_{PS}$ represents the mass scale of the new pseudoscalar interaction.
With the planned precision, PIENU will be sensitive to mass scales of $\Lambda_{PS} \sim O(1000 ~ {\rm TeV})$,  
which is far beyond the reach of collider experiments. Scalar couplings due to new physics can also induce changes 
to $R_{e / \mu}$ through higher order loop corrections \cite{r10}.

\section{The Experimental Setup}
\label{sec-1}
\begin{figure}[!t]
\includegraphics[trim=0cm 0cm 0cm 0cm, clip=true,angle=-90,width=6.0cm]{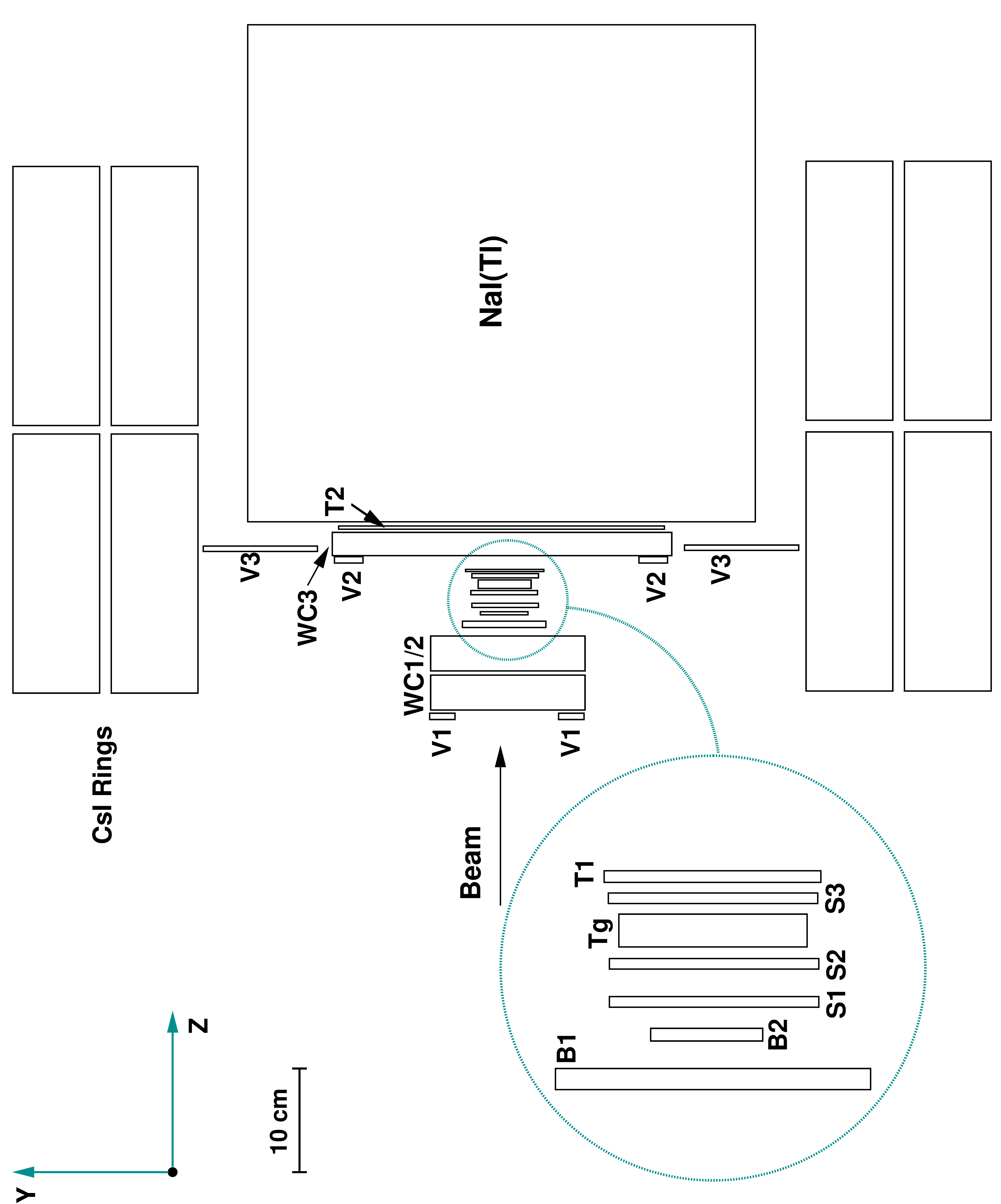}
\centering
\caption{Schematic view of the PIENU setup. The $\pi^{+}$ beam comes from the left and stops in the active target (Tg).
Time of the incoming pion is measured by the B1 scintillator, while the time of the decay positrons is measured by the T1 scintillator. 
The energy is measured by the NaI(Tl) and CsI calorimeters.}
\label{fig-1}  
\end{figure}
\begin{figure}[!t]
\includegraphics[trim=0cm 0cm 0cm 0cm, clip=true,width=7.0cm]{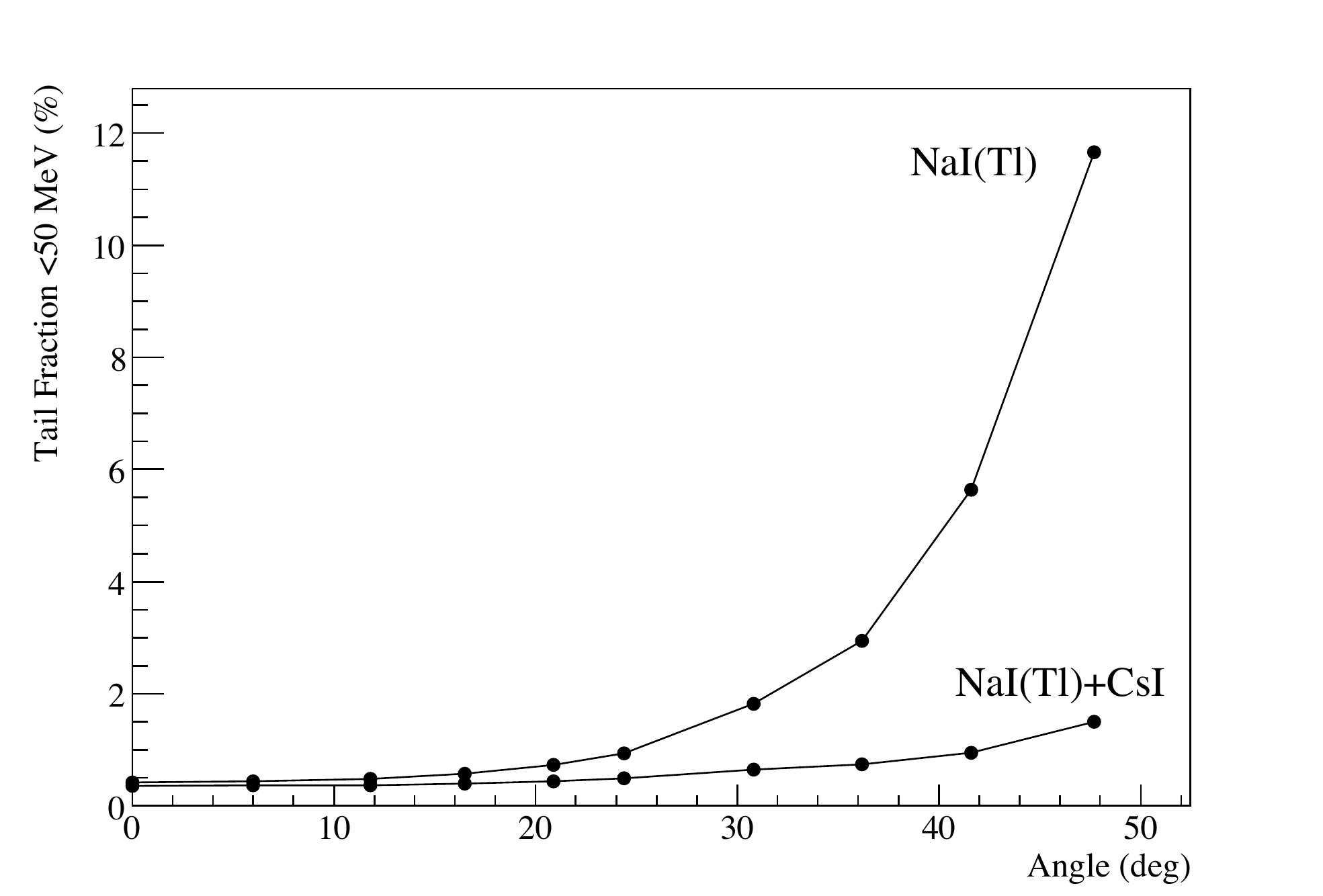}
\centering
\caption{Fraction of events with energy below 50 MeV in the NaI(Tl) response to a 70 MeV positron beam as
a function of its angle with respect of the crystal. The addition of the energy measured by the CsI array 
contains the low energy tail within about 2\%.}
\label{fig-2}  
\end{figure}

The PIENU experiment will obtain $R_{e / \mu}$ from the ratio of positron yields from the  $\pi^{+}\rightarrow e^{+}\nu$ decay
($E_{e^+}=69.3$ MeV) and the $\pi^{+}\rightarrow \mu^{+}\nu$ decay followed by the 
$\mu^{+} \rightarrow e^{+}\nu\bar{\nu}$ decay ($\pi^{+}\rightarrow\mu^{+}\rightarrow e^{+}$, $E_{e^+}=0-52.3$ MeV).
By measuring the decay positrons from both decays at the same time with the same apparatus (see fig.~\ref{fig-1}), many 
normalization factors like the solid angle cancel at the first order and only small energy-dependent
effects have to be taken into account as correction.
The experiment is based on a stopped beam technique: a 75 MeV/c $\pi^{+}$ beam impinges on an active scintillator target
where it stops via energy loss (50-100k/s pion stop rate). 
The high-purity pion beam produced with an energy-loss separation technique \cite{r4}, has a very low ($\le2\%$) positron content.
Pions decay at rest in the target, which is thick enough to contain
the muons from the $\pi^{+}\rightarrow \mu^{+}\nu$ decay. 
The energy of the positrons from the target is measured with a large (48$\times$48 cm) NaI(Tl) 
crystal with 25\% solid angle and about 1\% FWHM energy resolution at 70 MeV.
The NaI(Tl) is surrounded by 97 CsI crystals for shower leakage detection (fig.~\ref{fig-2}).
Identification of pions is achieved with scintillators, wire chambers and silicon microstrip 
detectors before the target. Silicon detectors are important for identifying pion decays in flight 
before the target. After the target other scintillators and silicon detectors are used for tracking and identification
of decay positrons. Another scintillator and wire chambers are placed in front of the NaI(Tl) calorimeter
for defining the positron acceptance.
The scintillators before and after the target provide the measurement of the decay time.

\section{Branching Ratio Measurement}
\label{sec-2}
\begin{figure}
\includegraphics[trim=0cm 0cm 0cm 0cm, clip=true,width=12.0cm]{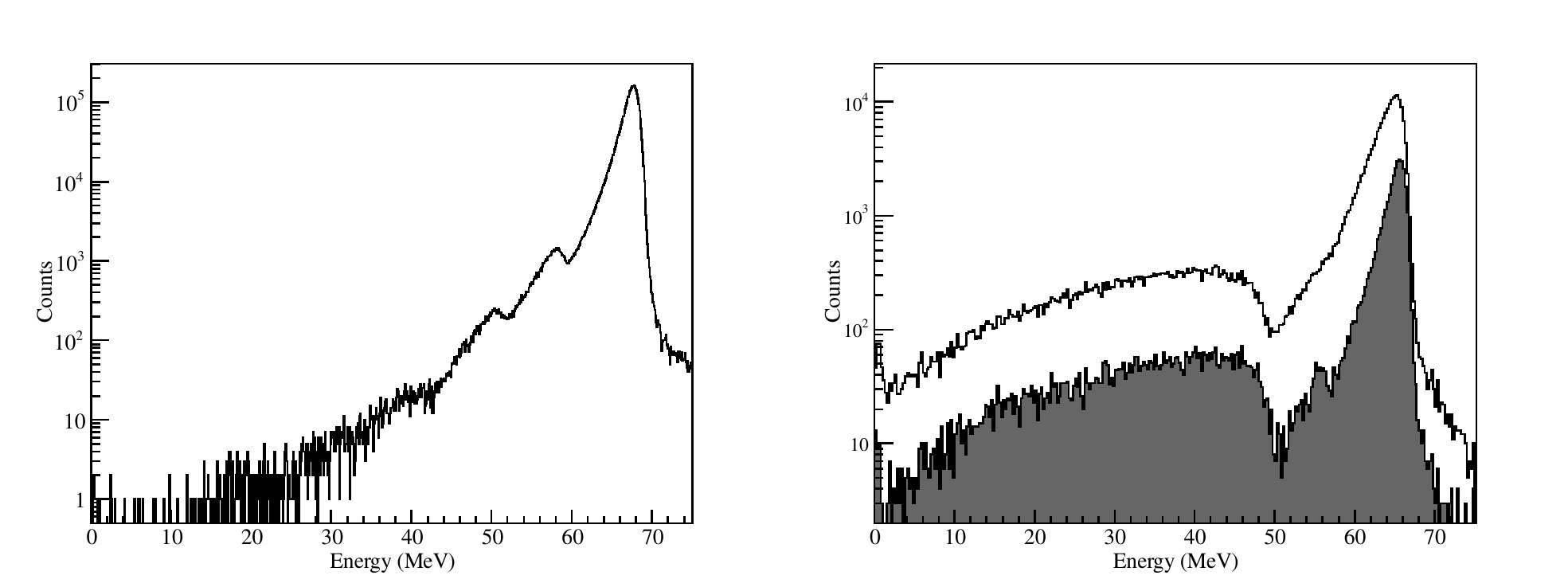}
\centering
\caption{(Left) NaI(Tl) response to a positron beam. (Right)  Measured   
$\pi^{+}\rightarrow \mu^{+} \rightarrow e^{+}$ + $\pi^{+}\rightarrow e^{+}\nu$ energy spectrum in NaI(Tl) with
selection cuts for suppressing the $\pi^{+}\rightarrow \mu^{+} \rightarrow e^{+}$ component with two different
acceptance cuts on the radius of the wire chamber WC3 (white (R=60mm) and shaded (R=30mm) histograms).}
\label{fig-3}  
\end{figure}

The analysis strategy is based on dividing the positron energy spectrum into low-energy (containing mostly
$\pi^{+}\rightarrow \mu^{+} \rightarrow e^{+}$ decays) and high-energy ($\pi^{+}\rightarrow e^{+}\nu$ decays) regions.
The regions are divided slightly above the end-point of the $\pi^{+}\rightarrow \mu^{+} \rightarrow e^{+}$ 
spectrum (the ``Michel edge''). The energy-based separation allows to identify the two decays, 
but backgrounds still remain to be identified and subtracted. 
The high energy region contains some fraction of the 
$\pi^{+}\rightarrow \mu^{+} \rightarrow e^{+}$ decays due to charged and neutral pileup effects.
The low-energy region contains a tail of $\pi^{+}\rightarrow e^{+}\nu$ decays. 
The tail originates from energy leakage
from the NaI(Tl) and radiative decays ($\pi^{+}\rightarrow e^{+}\nu\gamma$). The low-energy tail
is buried under the large amount of $\pi^{+}\rightarrow \mu^{+} \rightarrow e^{+}$ events and
its knowledge represented the major uncertainty of previous experiments.
In fig.~\ref{fig-2}, the low energy tail fraction of the calorimeter response to a positron beam is showed as
a function of the positron angle: the addition of the leaking energy measured by the CsI array successfully
contains the tail within 2\% level. The positron beam data, together with the $\pi^{+}\rightarrow e^{+}\nu$
data and detailed simulations will be used to extract the tail with high accuracy.
While investigating the NaI(Tl) response, photonuclear effects taking place in the crystal were detected
for the first time \cite{phn}: photons from the electromagnetic shower can be captured by Iodine nuclei with
subsequent neutron emission. If neutrons escape detection, peaks are visible on the tail of the crystal response (see fig.~\ref{fig-3}).
The branching ratio is obtained from a simultaneous fit of the time distributions of the low- and high-energy
events, taking into account signal and background distributions.
The final result is obtained applying the low-energy tail correction and other smaller corrections due
to acceptance and muon decays in flight in the target.
\begin{figure}
\includegraphics[trim=0cm 0cm 0cm 0cm, clip=true,width=6.0cm]{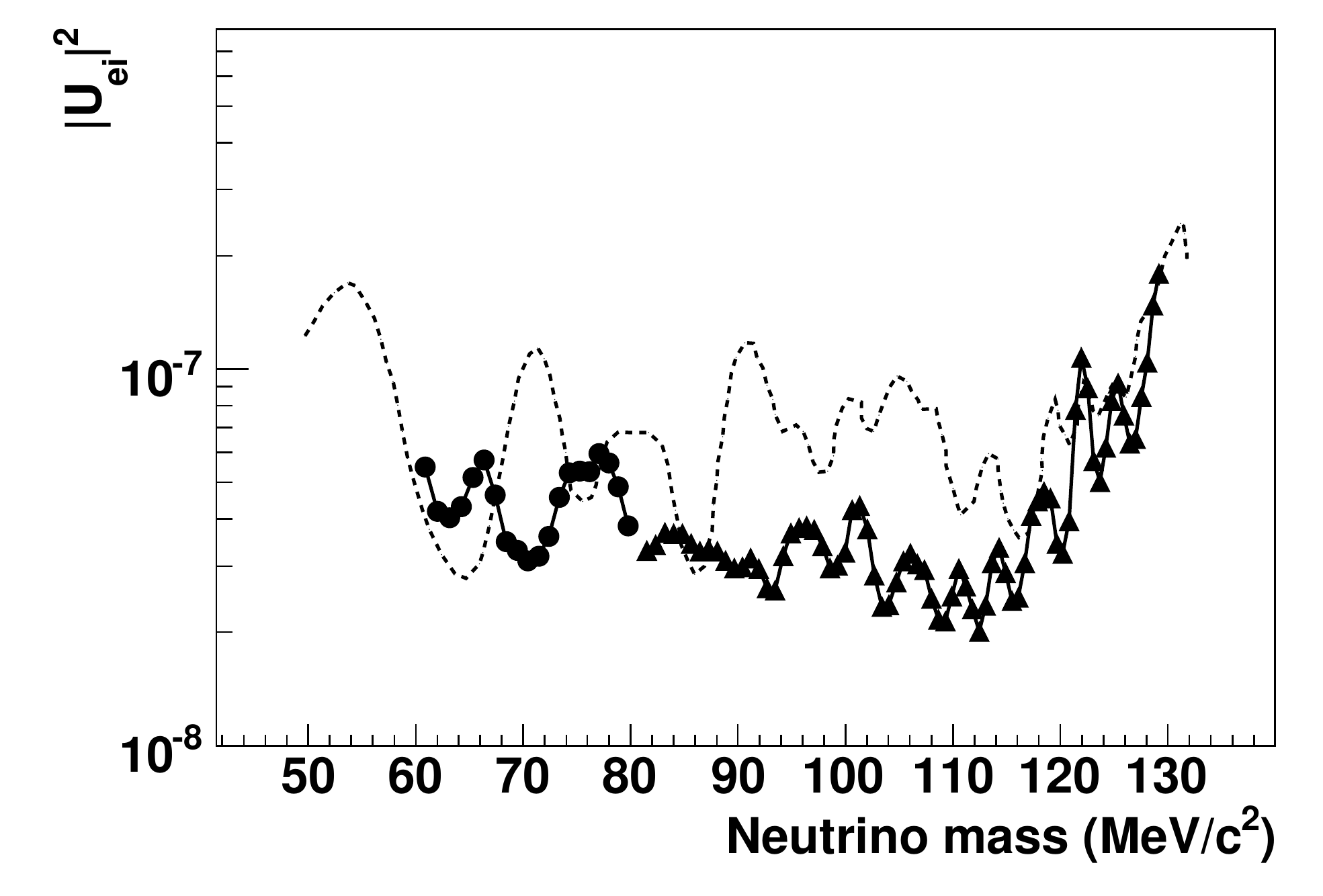}
\centering
\caption{Upper limit to the neutrino mixing matrix element. The dashed line
is the result of a previous experiment. The black points are the latest TRIUMF PIENU result.}
\label{fig-4}  
\end{figure}
\section{Summary}
\label{sec-3}
The PIENU experiment aims at the most precise measurement of the $R_{e / \mu}$ branching ratio
which will also provide the most stringent test of electroweak universality.
Technical papers have been published or are in preparation, while a first physics result 
on a massive neutrino search (fig.~\ref{fig-4}) has been published \cite{nu}.
Standard neutrinos can mix with hypothetical sterile neutrinos $\nu_{h}$. Such new states
might be seen considering the $\pi^{+}\rightarrow e^{+}\nu$ decay at rest in the PIENU experiment.
Since the kinematics is completely fixed, the presence of a massive neutrino will be detected 
as an additional peak in the energy spectrum. 
Concerning $R_{e / \mu}$, a portion ($\approx$ 1/10) of the data is being analyzed and results will be soon available,
providing a result already competitive with the old experiments.
In a second phase, the full dataset will be considered for the final result.

\end{document}